\documentclass[aps,pra,twocolumn,groupedaddress,floatfix,nofootinbib,longbibliography]{revtex4-2}
\usepackage{amsmath}
\usepackage{longtable}
\usepackage{amssymb}
\usepackage{graphicx}
\usepackage{mathrsfs}
\usepackage{ntheorem}
\usepackage{mathtools}
\usepackage{enumerate}
\usepackage{enumitem}
\usepackage{physics}
\usepackage{subfigure}
\usepackage{overpic}
\usepackage{epstopdf}
\usepackage{array}
 \usepackage{multirow}
 \usepackage{array}

\usepackage[most]{tcolorbox}
\usepackage{bm}
\usepackage{color,soul}
\usepackage[bookmarks=false]{hyperref}
\usepackage{xcolor}
\hypersetup{colorlinks=true,citecolor=blue,
linkcolor=blue,urlcolor=blue,pdfstartview=FitH,
bookmarksopen=true}
\usepackage{booktabs}
\newcommand{\ot}{\otimes}

\newcommand{\bkt}[2]{\langle{#1}|{#2}\rangle}
\def\blue#1{\textcolor{black}{#1}}
\def\red#1{\textcolor{black}{#1}}

\def\redhjh#1{\textcolor{black}{#1}}
\def\blueH#1{\textcolor{black}{#1}}
\begin{document}
\title{{Mitigating Source and Detection Noises in Auto-correlative Weak-Value Amplification}}

\author{Xiang-Yun Hu$^{1}$} 
\email{Contact author: xyhu@cug.edu.cn}
\author{Jing-Hui Huang$^{1,2}$} 
\email{Contact author: jinghuihuang@cug.edu.cn}
%\affiliation{Hubei Subsurface Multiscale Image Key Laboratory, School of Geophysics and Geomatics, China University of Geosciences, Lumo Road 388, 430074 Wuhan, China.}
\author{Fei-Fan He$^{4}$} 
\author{Guang-Jun Wang$^{5}$}
\author{Adetunmise C. Dada$^{3}$} 
\email{Contact author: Adetunmise.Dada@glasgow.ac.uk}
%\email{Contact author: Adetunmise.Dada@glasgow.ac.uk}
\address{$^{1}$Hubei Subsurface Multiscale Image Key Laboratory, School of Geophysics and Geomatics, China University of Geosciences, Lumo Road 388, 430074 Wuhan, China. }
%\author{Adetunmise C. Dada$^{3}$} %\email{Email:Adetunmise.Dada@glasgow.ac.uk}
%\author{Jeff. S. Lundeen$^{2}$} %\email{Email:jeff.lundeen@gmail.com}
%\author{Kyle M. Jordan$^{2}$} %\email{Email:kjordan@uottawa.ca}
%\author{Guang-Jun Wang$^{4}$} %\email{Email:gjwang@cug.edu.cn}
%\author{Xue-Ying Duan$^{4}$}
%\address{$^{1}$school of Geophysics and Geomatics, China University of Geosciences, Lumo Road 388, 430074 Wuhan, China. }
\address{$^{2}$Department of Physics and Centre for Research in Photonics, University of Ottawa, 25 Templeton Street, Ottawa, Ontario, Canada K1N 6N5 }
\address{$^{3}$School of Physics and Astronomy, University of Glasgow, Glasgow G12 8QQ, UK }
\address{$^{4}$ Institute of Optics and Electronics, Key Laboratory of Science and Technology on Space Optoelectronic Precision Measurement, Chinese Academy of Sciences, Chengdu 610209, China}
\address{$^{5}$ School of Automation, China University of Geosciences, Lumo Road 388, 430074 Wuhan, China.}

\begin{abstract}
Weak-value amplification (WVA), \blue{a post-selection-based technique that amplifies weak physical signals by preparing nearly orthogonal pre- and post-selected quantum states}, is intrinsically limited by various kinds of technical noise, which distorts amplified weak values, especially when discarding photons in post-selection. While prior work established the efficacy of auto-correlative weak-value amplification (AWVA) under Gaussian noise, practical implementations face challenges from band-limited laser-source noise and detection noise (including shot noise and electrical noise). Here, we demonstrate that the AWVA protocol robustly suppresses both laser-power fluctuations and detection noise. Numerical experiments in Simulink further reveal AWVA’s dual advantage: under high-power conditions, the noise-reduction superiority of AWVA over WVA becomes increasingly pronounced as input laser power increases, whereas in detection-limited regimes AWVA achieves an order-of-magnitude lower uncertainty, closely approaching the Cram\'er--Rao bound. Crucially, this work demonstrates that AWVA improves precision in both high-power (laser-noise-dominated) and photon-starved (detector-noise-dominated) regimes, thereby bridging these operating extremes and advancing precision in applications from gravitational-wave detection to hybrid quantum systems.

\end{abstract}
\maketitle

\section{Introduction}
\label{intro}
Environmental noise fundamentally limits the precision of quantum measurements~\cite{PhysRevA.88.013606,Sza_kowski_2017,MA2021127027,Abe2016,Ban2017}, from probing nanoscale spin dynamics to detecting gravitational waves~\cite{Abbott_2016,PhysRevD.103.044006,Cornish_2015,PhysRevD.105.024066,PhysRevD.104.063034}. While weak-value amplification (WVA)~\cite{AAV} enhances parameter estimation by amplifying small signals through post-selection, its sensitivity to noise—particularly non-Gaussian and spectrally structured disturbances—remains a critical challenge~\cite{RevModPhys.86.307,PhysRevA.102.042601,Yin2021,PhysRevLett.126.020502,PhysRevLett.126.220801,PhysRevLett.126.100403,Xia2023,apxr.202400136}.
This limitation is acute in real-world scenarios, such as optomechanical sensing~\cite{RevModPhys.73.357} and gravitational-wave interferometry~\cite{Abbott_2016}, where band-limited laser source noise and detection noise dominate.

Weak-value amplification leverages post-selection between nearly orthogonal quantum states to generate anomalously large weak values, enabling parameter estimation beyond the bounds of projective measurement~\cite{RevModPhys.86.307}. This counterintuitive feature has revolutionized ultrasensitive metrology, achieving
transverse optical deflections~\cite{PhysRevLett.102.173601}, 
velocities~\cite{2013Weak,PhysRevA.105.013718},  
angular rotation shifts~\cite{DELIMABERNARDO20142029,PhysRevLett.112.200401}, 
single-photon nonlinearity~\cite{PhysRevLett.107.133603,Hallaji2017},
temperature~\cite{10.1063/1.5027117,Li:19},
the weak magnetic fields detection~\cite{Huang:22},
photonic spin hall effect~\cite{doi:10.1126/science.1152697,PhysRevA.84.043806,PhysRevApplied.13.014057,10.1063/5.0184336}, 
frequency shifts~\cite{PhysRevLett.111.033604,10.1109/JPHOT.2019.2942718}.
Although \blue{postselection discards most photons}, an increasing number of experiments and theories have demonstrated the WVA technique~\cite{PhysRevX.4.011031,PhysRevA.80.041803,PhysRevA.103.053518,PhysRevLett.118.070802} and its modifications~\cite{PhysRevLett.116.100803,Yin2021,Song2021,PhysRevA.105.013718,PhysRevLett.128.040503,PhysRevA.105.033521,PhysRevA.97.063853,PhysRevA.97.063853,PhysRevA.106.012608,PhysRevA.109.053512} \blue{outperform conventional interferometers under specific technical noise conditions}.  
In particular, Amir Feizpour \textit{et al.} demonstrated that WVA can achieve higher SNRs in the presence of low-frequency noise (e.g.\ 1/\emph{f} noise) when amplifying single-photon \blue{nonlinearity}~\cite{PhysRevLett.107.133603}.
A recent experiment~\cite{PhysRevLett.134.080802} showed that implementing WVA inside an interferometer can enhance precision, with the signal-to-noise ratio (SNR) approaching the Cram\'er--Rao bound (CRB).  
The CRB sets the minimum achievable standard deviation for estimating an unknown parameter, as determined by the statistics of the measured pointer~\cite{PhysRevLett.115.120401,PhysRevX.4.011032,PhysRevX.4.011031,PhysRevA.106.022619,PhysRevA.107.042601,PhysRevA.105.013718}.

\begin{figure*}[htp!]
	\centering
\subfigure
{
	\vspace{-0.35cm}
	\centering
	\centerline{\includegraphics[scale=0.99,angle=0]{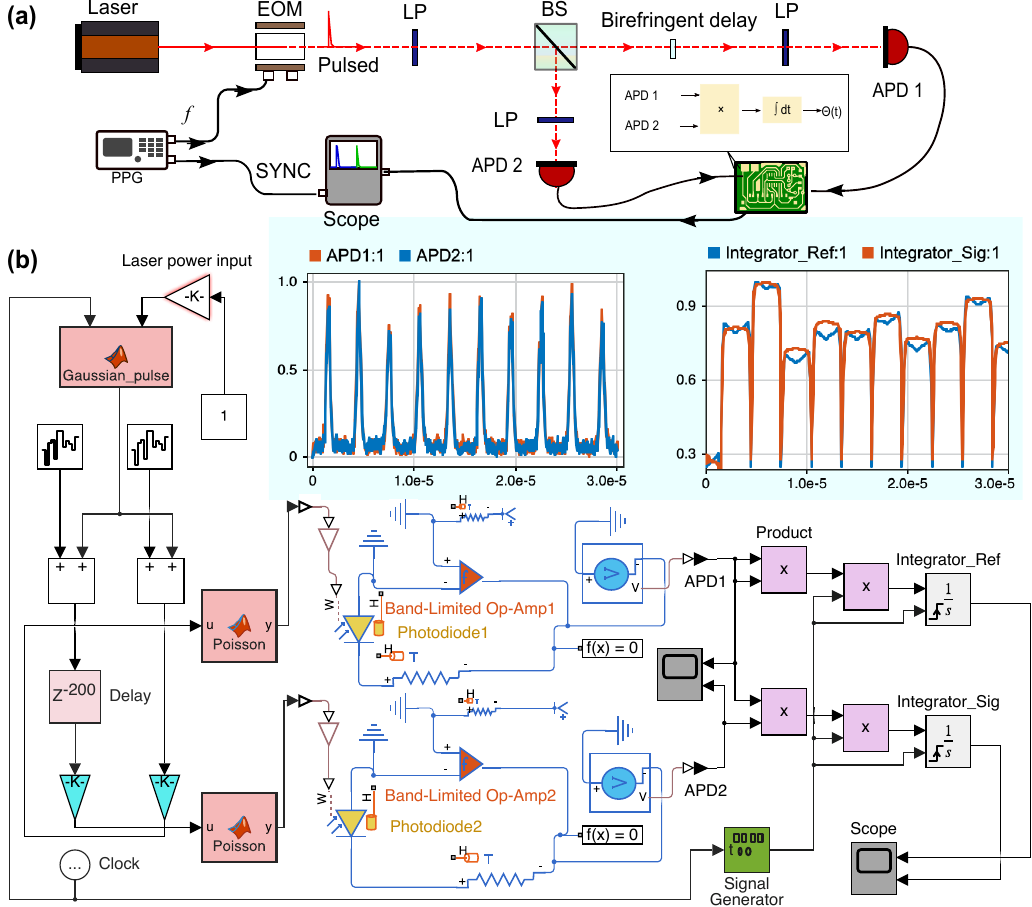}}
		\vspace{-0.35cm}
}
\vspace*{0mm} \caption{\label{Fig:Schemes_model2}
\textbf{Weak‐value amplification for birefringent time‐delay measurement under realistic noise.}  
(a) \blue{AWVA experimental setup}. A Gaussian pulse generated by a laser–EOM pair driven by a programmable pulse generator (PPG) is linearly polarised at $45^{\circ}$. A beam splitter (BS) creates two paths. Path 1 acquires a weak birefringent delay, is post-selected by a polariser at $(\alpha-45^{\circ})$, and is detected by APD1. Path 2, with the same polariser angle, serves as a reference detected by APD2. The two photodiode outputs are multiplied and integrated to obtain $\Theta(t)$.  
(b) {Simulink model schematic showing band-limited laser noise $N_s$ added at the source and detector noise $N_d$ (shot + electronic) added at each APD. The Simulink model is available in Ref.~\cite{DVN/0WERLH_2025}.
}
}
%\caption{\label{Fig:Schemes_model2} \red{  \textbf{Weak value amplification of measuring birefringent time delay under various noises.} (a) The AWVA protocal. The Gaussian pulse is produced by the laser and electro-optic modulator (EOM). Where the EOM is controlled by the programmable single pulse generator (PPG).  Then photons are pre-selected by a linear polarizer (LP) with an optical axis set at $45^\circ$.  Then the beam is divided by a beam splitter (BS).  Then, the photons are post-selected with an optical axis set at $\alpha - 45^\circ$, and are measured with an amplified photodiode (APD1).  The optical axis of LP3 is also set at $\alpha - 45^\circ$, and photon arrival times are measured with APD2. Finally, the detected signals at APD1 and APD2 passed through the product module ($\times$) and the integral module ($\int dt$) to calculate ${\rm \Theta}(t)$.   (b) The simulation model on Simulink. The band-limited source noise and the detection noise (including shot noise and electrical noise) are simulated.   }  }
\end{figure*}

Auto-correlative weak-value amplification (AWVA) suppresses Gaussian noise by correlating time-delayed measurement outcomes, as demonstrated in idealized settings~\cite{AWVA}. 
However, practical quantum and interferometric systems face non-Gaussian noise with distinct spectral and physical origins—band-limited laser-source noise and various detection noise that limit real-world applicability~\cite{PhysRevA.38.5938,PhysRevD.45.2843,WU2020124253,PhysRevE.101.052205}. {Whereas Ref.~\cite{AWVA} considered only additive Gaussian noise, the present work extends AWVA to \emph{band-limited} laser noise and detector (shot/electronic) noise. %
Laser-source noise, driven by pump instability and thermal drift~\cite{Kim:16}, typically governs classical interferometry, whereas detection noise~\cite{Xiang:22} prevails in low-photon quantum regimes. Existing techniques fail to address these distinct yet pervasive noise mechanisms simultaneously.}

{
Here, we validate AWVA's universal advantage over standard WVA under realistic noise conditions. Through analytical modeling and controlled Simulink simulations, we demonstrate AWVA's dual capability: it suppresses both high-power laser noise and photon-starved detection noise while approaching the CRB in detection-noise-limited scenarios. Crucially, AWVA achieves this without prior knowledge of noise spectra or secondary filtering---a key advance for real-time sensing. By unifying noise-robust correlation strategies across classical and quantum domains, our results establish AWVA as a versatile tool for optical gyroscopes, gravitational-wave detection, and quantum sensing.
}

\section{The AWVA protocol on Simulink}
Fig.~\ref{Fig:Schemes_model2}(a) illustrates the AWVA protocol setup~\cite{AWVA} for measuring a birefringent delay. In contrast, standard WVA (described in Ref.~\cite{PhysRevLett.105.010405}) omits the 50:50 beam splitter and reference arm. The key difference is that AWVA adds an extra light path (reference arm) parallel to the measurement arm, enabling an auto-correlation of outputs.
%The AWVA technique includes an initial preparation of the measured system  $\ket{\Phi_{i}} $ and the pointer $\ket{\Psi_{i}}$, weak interaction between the system and the pointer, a post-selection on the system \red{via projection on} state $\ket{\Phi_{f}}$ and a projective measurement on the pointer $\ket{\Psi_{f1}}$, the same projective measurement on the pointer $\ket{\Psi_{f2}}$ without the weak interaction\blue{. Finally, determination of} the auto-correlative intensity $\rm \Theta$ between the outcomes of the pointers $|\bkt{\Phi_{f1}}{\Phi_{i}}|^{2}$ and $|\bkt{\Phi_{f2}}{\Phi_{i}}|^{2}$. 
%
The AWVA technique includes an initial preparation of the measured system  $\ket{\Phi_{i}}$ and the pointer $\ket{\Psi_{i}}$, followed by a weak interaction between the two. 
\blue{The system is then post-selected via projection onto the state $\ket{\Phi_{f}}$, and the pointer is projectively measured to yield $\ket{\Psi_{f1}}$.  
A second, identical projective measurement is applied to the pointer state $\ket{\Psi_{f2}}$ obtained without the weak interaction.  
Finally, the auto-correlative intensity $\Theta$ is evaluated from the two outcomes} $|\bkt{\Phi_{f1}}{\Phi_{i}}|^{2}$ and $|\bkt{\Phi_{f2}}{\Phi_{i}}|^{2}$.

The initial pointer with the Gaussian profile is prepared in the time domain:
\begin{eqnarray}
\label{Eq:initial-pointer}
I^{in}(t) &=& \blueH{ P^{in} \left| \left\langle {t} | \Psi_{i}\right\rangle\right|^{2}} \nonumber \\
&=&P^{in} \frac{1} {(2 \pi \zeta^{2})^{1/4}}  e^{-(t-t_{0})^{2}/4\zeta^{2}} \,.
\end{eqnarray}
where $P^{in} $ represents the pulse amplitude (units of W), and $\zeta$ is the pointer spread. 
Then, the system and the pointer are weakly coupled with the interaction $\hat{H}=\tau \hat{A}\ot \hat{p}$. The observable operator satisfies $\hat{A}=\ket{H} \bra{H}-\ket{V} \bra{V} $, and $\hat{p}$ is the momentum operator conjugate to the position operator $\hat{q}$~\cite{AAV,RevModPhys.86.307,PhysRevLett.111.033604,PhysRevLett.126.220801}. Here, $\ket{H}$ and $\ket{V}$ represent the horizontally and vertically polarized states, respectively. 
%Note that AWVA needs two outcomes from two auto-correlative channels, one of the signals comes from the weak measurement for detecting shift $\tau$ and we call it the \red{{``}measurement channel"}. Correspondingly, another signal retains the information of the initial pointer without the weak interaction and \red{did not} change with the time shift $\tau$, and we call it \red{{``}reference channel"}.
In the standard WVA protocol, the final state of the pointer in the measurement channel is given by:
\begin{eqnarray}
\label{inter_peobe_final}
\ket{\Psi_{f1}} 
&=&\bra{\Phi_{f}}e^{-i\tau\hat{A}\ot \hat{p}} \ket{\Psi_{i}}  \ket{\Phi_{i}} \nonumber \\
&\approx&\bra{\Phi_{f}}\left[ 1-i\tau\hat{A}\ot \hat{p}\right]\ket{\Psi_{i}}  \ket{\Phi_{i}}  \\
%&=&\bkt{\Phi_{f}}{\Phi_{i}}\left[ 1-i\tau A_{w}\hat{p}\right]\ket{\Psi_{i}} \nonumber \\
&=&\bkt{\Phi_{f}}{\Phi_{i}}e^{-i\tau A_{w}\hat{p}}\ket{\Psi_{i}}\, , \nonumber
\end{eqnarray}
where $A_{w}:={\bra{\Phi_{f}}\hat{A}\ket{\Phi_{i}}}/{\bkt{\Phi_{f}}{\Phi_{i}}}$ is the so-called weak value~\cite{AAV}. Note that the \red{approximation} in Eq.~(\ref{inter_peobe_final}) only holds \red{in the weak-measurement region}, where the time shift $\tau$ is much smaller than the pointer spread $\zeta$ (see values in Table.~\ref{Tab_parameters}). 
%Normally, the weak value $A_{w}$ is a complex number \cite{PhysRevA.76.044103}, \redk{whose real part is associated with the pointer shift in position space, while its imaginary part is associated with the pointer shift in momentum space.}
In this paper, the time shift $\tau$ in the position of the pointer is amplified by the real part of $A_{w}$:
$
\Delta\langle\hat{q}\rangle=\tau {\rm Re} \,\rm [A_{w}] \,
$~\cite{PhysRevA.85.052110}.
To amplify $\tau$, the system is pre- and post-selected into the states:
\begin{eqnarray}
\label{Eq:post-sel-sys}
\ket{\Phi_{i}}&=&{\rm sin} \bigg(\frac{\pi}{4}\bigg) \ket{H}+ {\rm cos} \bigg(\frac{\pi}{4} \bigg)\ket{V}, \\
\ket{\Phi_{f}}&=& {\rm sin} \bigg(-\frac{\pi}{4}+\alpha\bigg) \ket{H}+ {\rm cos} \bigg(-\frac{\pi}{4}+\alpha \bigg)\ket{V},
\end{eqnarray}
where $\alpha$ is the postselection angle. One then obtains the weak value as:
$
A_{w}=
%\frac{{\rm sin} (-\frac{\pi}{4}+\alpha) - {\rm cos} (\frac{\pi}{4}+\alpha)}{{\rm sin} (-\frac{\pi}{4}+\alpha) + {\rm cos} (\frac{\pi}{4}+\alpha)} =
-{\rm cot} \alpha \, ,
$
and the corresponding time shift $\tau$ can be obtained from the peak shift $\delta t=|\tau {\rm Re} A_{w}|= \tau  {\rm cot} \alpha$ of the signal detected by an Amplified photodiode (APD1), with the detected signal $I_{1}^{out}$ calculated from Eq.~(\ref{inter_peobe_final}) as:
\begin{eqnarray}
\label{Eq:schme1inter_peobe_final}
I_{1}^{out}(t;\tau)&=& \blueH{P^{in}}  |\bkt{\Phi_{f}}{\Phi_{i}}|^{2} e^{-2i\tau A_{w}\hat{p}} \left|\left\langle q | \Psi_{i}\right\rangle\right|^{2} \nonumber \\
&\approx &   \frac{P^{in}}{2} \frac{({\rm sin}\alpha)^{2}} {(2 \pi \zeta^{2})^{1/4}}   e^{-(t-t_{0}-\delta t)^{2}/4\zeta^{2}} \,.
\end{eqnarray}

In the AWVA protocol, for the measurement in the reference channel without the weak interaction, the signal detected by APD2 can be calculated by
\begin{eqnarray}
\label{Eq:schme1inter_peobe_final_reference}
I_{2}^{out}(t;\tau=0)&=& \blueH{P^{in}} |\bkt{\Phi_{f}}{\Phi_{i}}|^{2}  \left|\left\langle q | \Psi_{i}\right\rangle\right|^{2} \nonumber \\
&\approx &   \frac{P^{in}}{2} \frac{({\rm sin}\alpha)^{2}} {(2 \pi \zeta^{2})^{1/4}}   e^{-(t-t_{0})^{2}/4\zeta^{2}} \,.
\end{eqnarray}
\blue{Following} the previous work~\cite{AWVA}, ${\rm \Theta}_{}(\tau)$ is defined as
\begin{eqnarray}
\label{Eq:ACIdefine+noise}
&&{\rm \Theta}(t;\tau)=\int_{0}^{t}  I_{1}^{out}(t^{\prime};\tau) \times  I_{2}^{out}(t^{\prime})  dt^{\prime}\,.
\end{eqnarray}

{
In the previous work~\cite{AWVA}, we simulated the AWVA measurements under Gaussian noises with a simple assumption that the effect from all sources of noise is equivalent to adding a certain {type} of noise on APDs. Where the noises were {assumed to} be added directly to the final measured signal $I_{1}^{out}(t;\tau)$ at APD1 and $I_{2}^{out}(t;\tau)$ at APD2. 
While effective for benchmarking, this approach conflates the distinct physical origins and spectral profiles of laser source noise and detection noise—a simplification that limits applicability to real-world quantum metrology.
Thus, we rigorously separate laser source noise and detection noise in the measurement chain (Fig.~\ref{Fig:Schemes_model2}b). 
By modeling these noise regimes in Simulink, we will demonstrate AWVA’s ability to suppress both laser source noise and detection noise.
}

\section{Simulation and Results}

{
We quantify the precision of WVA and AWVA under dominant noise regimes using a Simulink model (Fig.~\ref{Fig:Schemes_model2}b). 
Two noise sources are explicitly modeled:
\textbf{Band-limited laser source noise} ($N_s$), added to the signal path with power $P_s$;
\textbf{Detection noise}, modeled as Poisson-distributed pre-detection photon-counting noise (shot noise), and electrical noise generated via photodiode and Op-Amp components to mimic electronic readout noise.
{In Fig.~\ref{Fig:Schemes_model2}, shot noise is modeled via a function block implementing Matlab's poissrnd() with incident power as input, while electrical noise is characterized by the parameters in Table~\ref{Tab_parameters}.}

\begin{table}[b]
\caption{\label{Tab_parameters}
\textbf{Simulation parameters in Simulink.} 
{Scenario~I and Scenario~II are simulated with the same parameters, except  $P_s$(S. I) and $P_s$(S. II) of the laser noise power. Note that the detection noise $N_d$, including the shot noise and electrical noise,  can not be quantified, since the amplitude of $N_d$ depends on the signal.}
 }
\begin{tabular}{llll}
\toprule
                         Type  & Parameter        & Meaning                      & Value                              \\ \hline
\multirow{4}{*}{\rotatebox[origin=c]{90}{WVA}}      
& $\tau$         & Birefringent delay          & $1\times 10^{-10}$ $\rm s$                           \\
& $\alpha$         & Postselection angle          & 0.01 rad                           \\
 & ${\rm Re} A_{w}$ & Weak value                   & 100     \\
& $\lambda$& Wavelength                  & 633 nm 
 \\ \hline
\multirow{5}{*}{\rotatebox[origin=c]{90}{Laser}}   
& $P^{in} $        & Amplitude                    & $10^{-3}$-$10^{4}$ W               \\
& $P_s$(S. I)      & Noise power             & $2 \times 10^{-14}$ $\rm W^2$/Hz             \\
& $P_s$(S. II)      & Noise power             & $2 \times 10^{-22}$ $\rm W^2$/Hz             \\
& $\zeta$      & Spread                       & $2\times 10^{-7}$ $\rm s$             \\
                           & $T$              & \blue{Pulse period}                 & \blue{$3\times 10^{-6}$ s}                 \\
                           & $t_{0}$          & \blue{Time offset for pulse}  & \blue{$1.5 \times 10^{-6}$ s }             \\ \hline
\multirow{5}{*}{\rotatebox[origin=c]{90}{Detection}} & $1/f_d$          & Sample time                  & $5 \times 10^{-11}$ s              \\
                           & $S_d$            & Sensitivity                  & $113 \times 10^{-10}$ A$\rm m^2$/W \\
                           & $I_d$            & Dark current                 & $1 \times 10^{-6}$ A               \\
                           & $A_d$            & Gain                         & 10000                              \\
                           & $f_d$            & Bandwidth                    & \redhjh{$2 \times 10^{10}$} Hz     \\
\bottomrule
\end{tabular}
\end{table}
Precision is characterized by the standard deviation (Std) of the time-delay estimation for the WVA protocol ($\sigma_{1}$), the AWVA protocol (${\sigma}_{1}$), and the Cram{é}r-Rao bound (${\sigma}_{3}$). For AWVA, 
$\sigma_{2}$ derives from the peak change in ${\rm \Theta}(t;\tau)$
, while $\sigma_{1}$ for WVA uses intensity centroid shifts. The CRB  ${\sigma}_{3}$ sets the theoretical quantum limit~\cite{PhysRevLett.115.120401}. In this paper, we define
}
{
\begin{eqnarray}
\label{Eq:std_WVA}
{\sigma}_{1}=\frac{{\rm Std} [ {\rm Loc} (I_{1}^{out}(t;\tau))- {\rm Loc }(I_{2}^{out}(t;0\,{\rm s} )) ]}{ \langle {\rm Loc} (I_{1}^{out}(t;\tau))- {\rm Loc }(I_{2}^{out}(t;0\,{\rm s}) \rangle }  \times \tau,
\end{eqnarray}
\begin{eqnarray}
\label{Eq:std_AWVA}
{\sigma}_{2}=\frac{{\rm Std} [ {\rm Peak} ({\rm \Theta}(t;\tau))- {\rm Peak }({\rm \Theta}(t;0\,{\rm s} )) ]}{ \langle {\rm Peak} ({\rm \Theta}(t;\tau))- {\rm Peak }({\rm \Theta}(t;0\,{\rm s}) \rangle }  \times \tau,
\end{eqnarray}
\begin{eqnarray}
\label{Eq:std_CFI}
{\sigma}_{3}^{-2}=  \mathcal{N} \int_{0}^{T} I^{in}(t-\tau)  \bigg[ \frac{\partial}{\partial \tau} {\rm ln} {\,} I^{in}(t-\tau) \bigg]^{2} dt  ,
\end{eqnarray}
where $T$ is the pulse period, $c$ is the speed of light, $\lambda$ is the wavelength of the photons, $\mathcal{N}={\mathcal{G} P^{in}T\lambda}/{\hbar c}$ is the total number of input photons, and $\mathcal{G}=0.93$ is the correction factor for the Gaussian pulse. 
In this work, the CRB band is calculated, including shot noise (fundamental quantum noise) but excluding electronic readout noise with the approximation of the large-$N$ limit.
{The} function Loc() {returns the temporal centroid (first moment) of} $I_{1}^{out}(t;\tau)$, and {the}  function Peak() {returns the maximum (peak value) of} ${\rm \Theta}(t;\tau)$. 
In this work, we investigate two different scenarios where laser-source noise or detection noise dominates.
}
\begin{figure*}[t]
\begin{minipage}{0.48\linewidth}
	\centering
\includegraphics[width=0.99\textwidth]{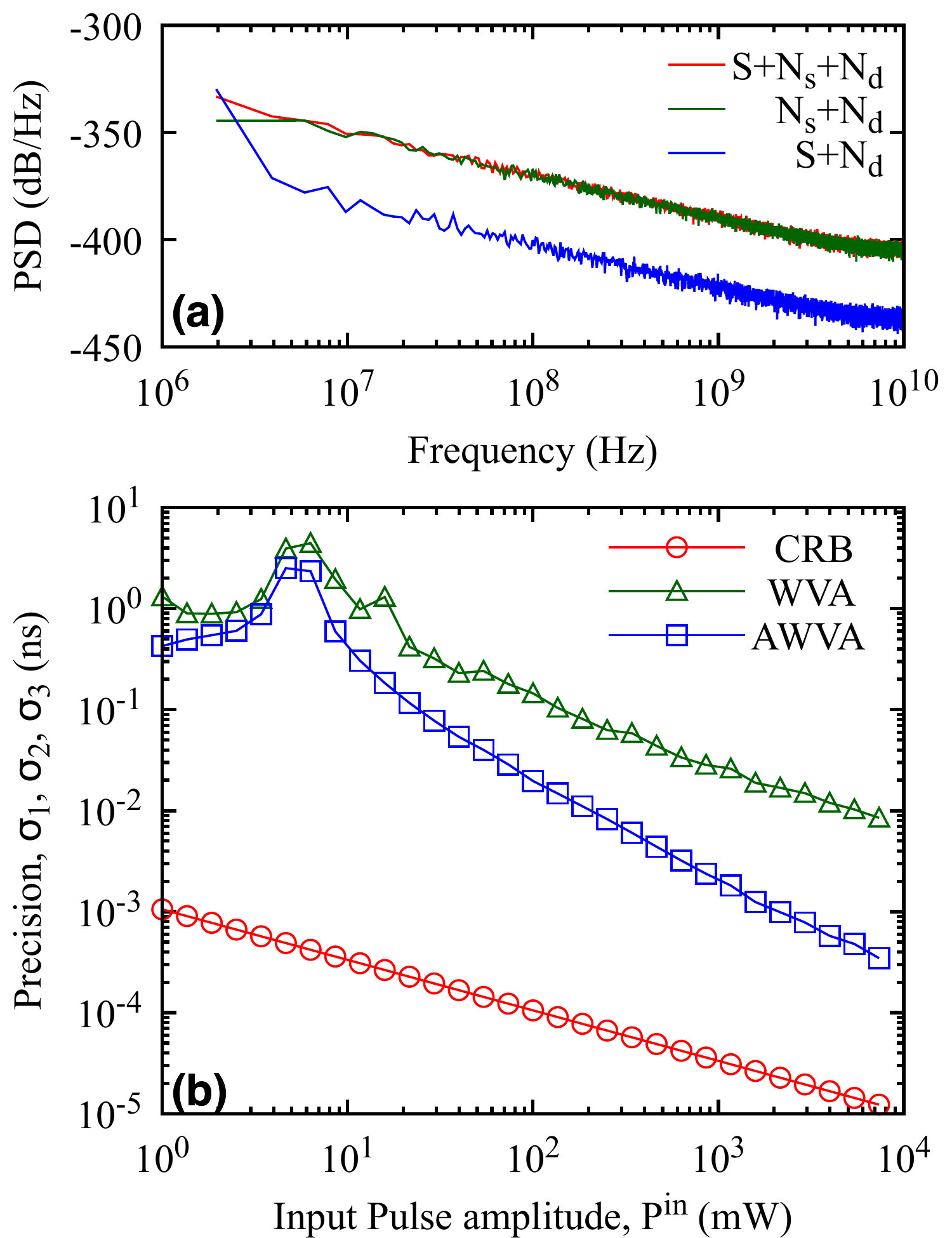}
\end{minipage}
\begin{minipage}{0.48\linewidth}
	\centering
\includegraphics[width=0.99\textwidth]{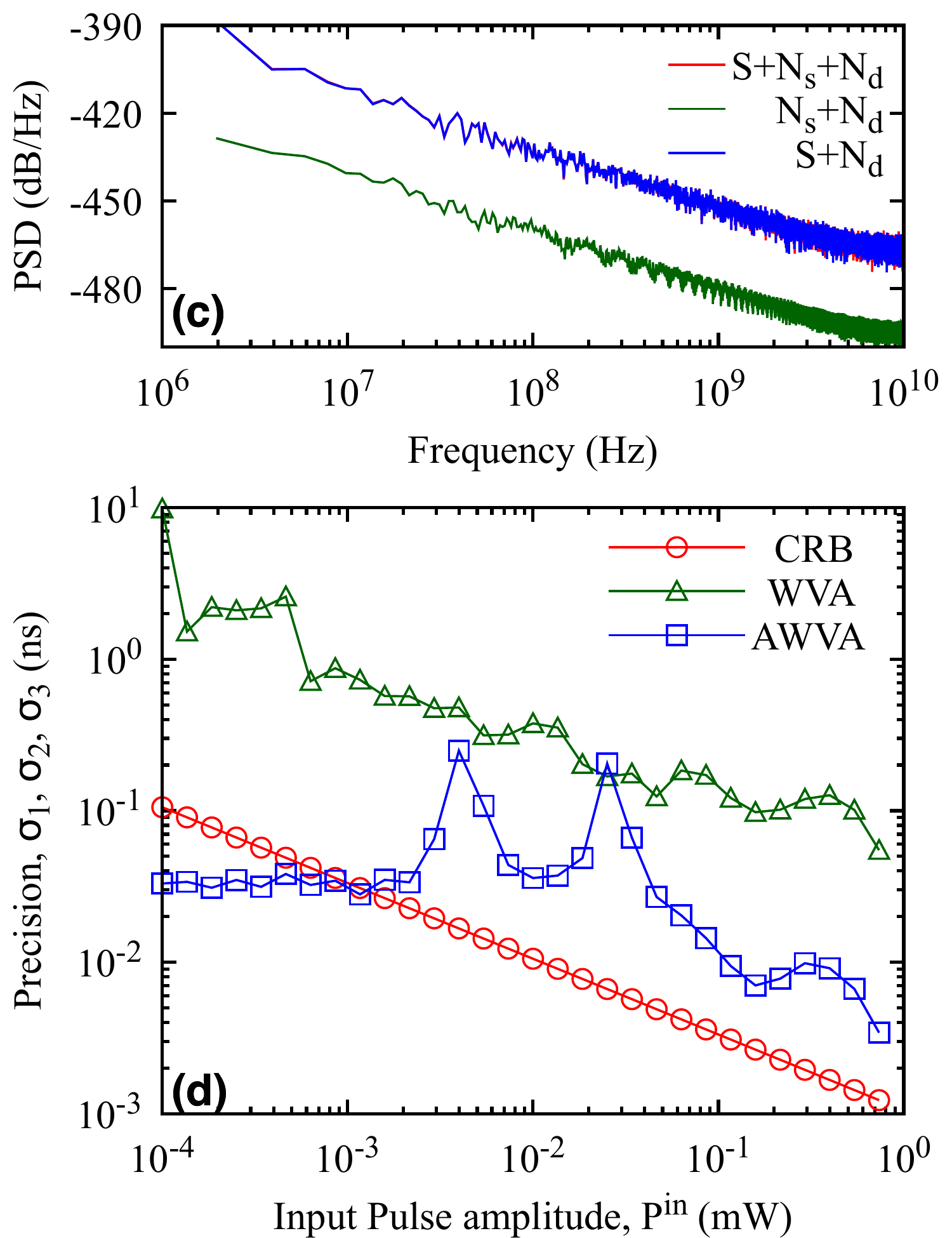}
\end{minipage}
\vspace*{-0mm} 
\caption{\label{Fig:Results}
\textbf{Simulated time delay measurement performance of WVA and AWVA protocols.} The value of the time delay is set at $\tau=0.1$ ns.
(a) Power spectral density (PSD) of WVA signals in Scenario I (high-power laser, $P^{in}=100$ mW). Laser source noise $N_s$ dominates over detection noise $N_d$. \redhjh{Here, the labels ``S+$N_s$+$N_d$", ``S+$N_d$", and ``$N_s$+$N_d$" denote distinct combinations of the ideal signal (S), laser source noise ($N_s$), and detection noise ($N_d$).
}
(b) Time delay estimation precision of the WVA, AWVA, and Cramér-Rao bound (CRB) limits for Scenario I. 
\redhjh{
Here, the CRB bound is calculated based on the formula (\ref{Eq:std_CFI}).
(c) PSD results for Scenario II (low-power, $P^{in}=0.01$ mW).
(d) Time delay estimation precision of the WVA, AWVA, and CRB limits for Scenario II.
}
 }
\end{figure*}

\blueH{
To ensure a fair comparison between the WVA and AWVA techniques, which use different observables (the intensity profile $I(t)$ and the autocorrelation ${\rm \Theta}(t)$, respectively), we define the estimation precision for the time delay $\tau$ itself. For any unbiased estimator $y$ that is linearly proportional to $\tau$ (i.e., $\langle y \rangle = k \tau$), the standard deviation of the estimated $\tau$ is given by:
\begin{equation}
\mathrm{Std}[\tau] = \frac{\mathrm{Std}[y]}{k}.
\label{eq:general_precision}
\end{equation}
The quantity $\mathrm{Std}[\tau]$ is a figure of merit for the estimation error that is independent of the specific choice of estimator $y$. 
For the WVA technique, we use $y = {\rm Loc} (I_{1}^{out}(t;\tau))- {\rm Loc }(I_{2}^{out}(t;0\,{\rm s} )) $ and calculate $k = |d\langle I(t) \rangle / d\tau|$ around the operating point. This leads directly to the definition of $\sigma_1$ in Eq.~\eqref{Eq:std_WVA}. 
Similarly, for the AWVA technique, we use $y = {\rm Peak} ({\rm \Theta}(t;\tau))- {\rm Peak }({\rm \Theta}(t;0\,{\rm s} )) $ and its corresponding slope to define $\sigma_2$ in Eq.~\eqref{Eq:std_AWVA}. Consequently, both $\sigma_1$ and $\sigma_2$ represent $\mathrm{Std}[\tau]$, allowing for their direct comparison.
}

{
\textbf{Scenario I (Laser noise dominates)}: Set high-power laser and larger laser source noise $N_s$. We show the power spectral density (PSD) results of three simulations for the WVA protocol in Fig.~\ref{Fig:Results}(a), including different signal combinations of the ideal signal input (S), the laser source noise $N_s$ and the detection noise $N_d$. When setting $P^{in}=100$ mW, the curve PSD(S+$N_d$) is much lower than the curves PSD(S+$N_s$+$N_d$) and PSD($N_s$+$N_d$), indicating that $N_d$ contributes little to the final signal $I_{1}^{out}(t;\tau)$ thus the laser source noise dominates.
}

{
\textbf{Scenario II (Detection noise dominates)}: Set low-power laser and smaller laser source noise $N_s$.
Similarly, the PSD curves with setting $P^{in}=0.01$ mW are present in Fig.~\ref{Fig:Results}(c). 
In contrast to Scenario I, the curve PSD(S+$N_d$) almost overlaps with the curve PSD(S+$N_s$+$N_d$), indicating that $N_s$ contributes little to the final signal $I_{1}^{out}(t;\tau)$, thus the detection noise dominates.
}

{
Our numerical experiments quantify AWVA’s advantage across two noise-dominated regimes. For each input pulse amplitude $P^{in}$, we simulated 100 independent measurements of time delay $\tau$= 0.1 ns over $t_{stop}=3\times 10^{-4}$ s.
We represent the main simulation parameters in Table~\ref{Tab_parameters} and code in Ref.~\cite{DVN/0WERLH_2025}. Figure~\ref{Fig:Results}(b) and (d) compare AWVA and WVA performance in Scenario I (laser noise-dominated) and Scenario II (detection noise-limited), respectively.
It can be found that the statistical standard deviation $\sigma_2$ of the AWVA measurement is smaller than $\sigma_1$ of the WVA measurement,  despite the laser input power. {At $P_{\rm in}=0.01$ mW (Scenario II), WVA yields $\sigma_1\approx0.3$ ns whereas AWVA yields $\sigma_2\approx0.03$ ns (Fig.~\ref{Fig:Results}(d)), nearly 10× improvement.}
}

{
In Scenario I, the noise-reduction advantages of AWVA over WVA will be highlighted when increasing the input laser power. {AWVA’s $\sigma_2$ is ~10\% lower than $\sigma_1$ at the highest powers (Fig.~\ref{Fig:Results}(b)).}
In addition, Fig.~\ref{Fig:Results}(d) for Scenario II shows that precision in AWVA is improved by approximately one order of magnitude of std compared to WVA, demonstrating the ability to approach the CRB limit.
}

\section{Discussion and Summary}
{
In Scenario I, AWVA reduces the statistical uncertainty $\sigma_2$ by up to $10\%$ compared to WVA’s $\sigma_1$
at high laser power $P^{in}>100$ mW, where laser power fluctuations dominate (Fig.~\ref{Fig:Results}(b)). This scaling arises because AWVA’s auto-correlative post-processing suppresses laser noise, which grows with input power. In contrast, WVA’s single-shot measurements fail to leverage these correlations, leaving it susceptible to noise amplification at elevated powers—a critical drawback in high-power interferometric systems.
}

{
Scenario II highlights AWVA’s quantum advantage: under detection noise (shot and amplifier noise), AWVA achieves a remarkable reduction in $\sigma_2$ over WVA, closely approaching the CRB (Fig.~\ref{Fig:Results}(d)). This aligns with our analytical models, where AWVA’s correlated sampling cancels detection-stage noise correlations, enabling sub-shot-noise sensitivity. The CRB proximity confirms AWVA’s superiority in photon-limited regimes, overcoming WVA’s trade-off between amplification and noise susceptibility.
}

\blueH{
The sub-CRB precision in Fig.~2(d) (low input-amplitude regime) arises from finite sampling effects and a benchmark mismatch, not a physical surpassing of the fundamental limit. 
The CRB is an asymptotic limit for unbiased estimators under a specified noise model. Our precision points in Fig.~2 are computed from $N=100$ Monte-Carlo trials, making it possible for the resulting finite-sample RMSE to fluctuate slightly below the asymptotic CRB at low input powers. Importantly, our CRB curve assumes an ideal broadband detector, whereas the Simulink model includes a band-limited, slew-rate-limited amplifier with mild nonlinearities, which reshape the noise spectrum. While electronics add noise overall, such filtering can reduce the variance captured by our estimator relative to an idealized CRB. For Fisher information computed with the detector’s actual transfer function and noise PSD, the bound would shift accordingly and the apparent sub-CRB dip disappears. Thus, no fundamental limit is violated.}

In addition, the pronounced oscillations of curves $\sigma_1(P^{in})$ and $\sigma_2(P^{in})$ in Scenario II arise from finite simulation time $t_{stop}=3\times 10^{-4}$ and the stochastic nature of detection noise. While longer simulations would smooth these artifacts, computational constraints limit practical runtime due to the model’s complexity.
Nonetheless, our Simulink framework—employing engineering-grade parameters—provides a scalable platform to study detection noise, bridging the gap between idealized theory~\cite{PhysRevX.4.011031,PhysRevA.38.5938,PhysRevD.45.2843,WU2020124253,PhysRevE.101.052205} and real-world optoelectronic systems.

{
In summary, we demonstrate that AWVA universally enhances phase estimation under realistic non-Gaussian noise. {Our numerical model confirms that AWVA outperforms WVA under both dominant laser noise and dominant detector noise.} 
These results unify AWVA’s utility across classical and quantum sensing paradigms. 
By suppressing both laser noise and detection noise without prior spectral knowledge, AWVA demonstrates the ability to approach the CRB limit in detection-noise-limited scenarios. 
This dual capability is critical for hybrid systems, such as optomechanical sensors or gyroscopes, where noise sources coexist~\cite{Zhao:22,Liu:22}. 
Furthermore, AWVA’s computational simplicity—requiring only post-measurement correlation—enables real-time operation, which can be implemented in an FPGA-based~\cite{9751594} high-speed optical sensor.
While these results are based on simulation, they suggest AWVA could be implemented in real sensors to enhance precision. 
%Future work could involve experimental validation of AWVA in an interferometer.
Our results establish AWVA as a versatile strategy for applications that demand both high-power interferometry and photon-starved quantum sensing, thereby unifying classical and quantum noise resilience under a single protocol.
}

\begin{acknowledgments}
%We are in debt to Jeff. S. Lundeen and Kyle M. Jordan at the University of Ottawa for stimulating discussions.
This study was supported by the National Natural Science Foundation of China (Grants No.~42327803, No.~42504048, and No.~42488201). 
J-H. Huang acknowledges support from the Hubei Provincial Natural Science Foundation of China (Grant No.~20250650025), the Fellowship Program of China National Postdoctoral Program for Innovative Talents under Grant Number BX20250161, and the CSC. 
A.C.D. acknowledges support from the EPSRC, Impact Acceleration Account (Grant No.~EP/R511705/1).
\\

\textbf{Data availability}—The data that support the findings of this paper are openly available~\cite{DVN/0WERLH_2025}.
\end{acknowledgments} 

\bibliography{reference}
%\section{Appendix}

\end{document}